\documentclass[12pt]{iopart}
\usepackage{epsfig}
\usepackage{iopams}
\begin{document}

\title[Dynamics of a nanomechanical resonator]
{Dynamics of a nanomechanical resonator coupled to a
superconducting single-electron transistor}

\author{M.P. Blencowe\dag, J. Imbers\ddag,  and A.D. Armour\ddag}
\address{\dag Department of Physics and Astronomy, 6127 Wilder
Laboratory, Dartmouth College, Hanover, NH 03755, USA\\
\ddag School of Physics and Astronomy, University of Nottingham,
Nottingham,\\ NG7 2RD, UK}
\ead{\mailto{miles.p.blencowe@dartmouth.edu},\\\mailto{ppxji@nottingham.ac.uk},
\\\mailto{andrew.armour@nottingham.ac.uk}}

\begin{abstract}

   We present an analysis of the
   dynamics of a nanomechanical resonator coupled to a superconducting
   single electron transistor (SSET) in the vicinity of
   the Josephson quasiparticle (JQP) and double Josephson
   quasiparticle (DJQP) resonances. For weak coupling and wide separation of
   dynamical timescales, we find that for either superconducting
   resonance the dynamics of the resonator is
   given by a Fokker-Planck equation, i.e., the SSET behaves
   effectively as an equilibrium heat bath, characterised by an effective
   temperature,
   which also damps the resonator and renormalizes its frequency.
   Depending on the gate and drain-source voltage bias
   points with respect to the
   superconducting resonance, the SSET can also give rise to an instability in
   the mechanical resonator marked by negative damping and
   temperature within the appropriate Fokker-Planck equation.
   Furthermore, sufficiently close to a resonance, we find that the
   Fokker-Planck description breaks down.
   We also point out that there is a close analogy between coupling a
   nanomechanical
   resonator to a SSET in the vicinity of the JQP resonance and Doppler cooling of atoms
   by means of
   lasers.

\end{abstract}
\submitto{NJP}
\maketitle

\section{Introduction}
Nanomechanical single-electron transistors in which a mechanical
resonator forms the voltage gate of the transistor constitute a new and
interesting class of nanoelectromechanical system. The idea of
coupling a nanomechanical resonator to the island of a single
electron transistor (SET) as a mechanically compliant voltage gate
was proposed as a way of measuring the displacement of a
mechanical resonator with high
precision~\cite{white,bw,zb,nems}, since the conductance properties
of the SET are extremely sensitive to the resonator motion. Indeed
such devices have recently been used to measure the displacement
of a nanomechanical resonator with almost quantum limited
precision~\cite{mset1,mset2}.

The sensitivity with which a SET can be used to measure the
position of a nanomechanical resonator is ultimately limited by
the back-action of the SET on the dynamics of the resonator. The
back-action of electrons moving through a {\it normal state} SET
gated by a nanomechanical resonator was studied
recently~\cite{mmh,armour,noise,blencowe} and it was shown that, in
the regime where the energy associated with the applied bias
voltage is much larger than the resonator energy quanta, the SET
electrons act on the nanomechanical resonator in a way which is
closely analogous to an equilibrium thermal bath. In fact, the
dynamics of the resonator can be described by a Fokker-Planck
equation for a damped harmonic oscillator in contact with a
thermal bath at a fixed temperature~\cite{blencowe}. Very similar
results were obtained explicitly for a resonator coupled to a
tunnel junction~\cite{mm,Clerk1} and it was also shown by
Clerk~\cite{Clerk2} that such behaviour is expected to be generic
within the regime of linear response.

In contrast to normal state SETs where the current arises solely
from electron tunnelling and cotunnelling
processes~\cite{schoeller}, superconducting SETs (SSETs) can
support a wide range of different electronic processes which
contribute to the current including tunnelling or cotunnelling of
quasiparticles, coherent tunnelling of Cooper pairs and even
Andreev
reflection~\cite{averin,nakamura,fitzgerald,choi,djqp1,djqp2}.
Furthermore, there exist a number of current resonances for
particular values of the drain-source and gate voltages of the
SSET where current is carried by a combination of different
processes occuring at the source and drain junctions in turn. The
best known (and most readily observed experimentally) current
resonances for the SSET are the Josephson quasiparticle (JQP)
and double Josephson quasiparticle (DJQP) cycles where transport
occurs via a combination of coherent, resonant tunnelling of
Cooper pairs and incoherent quasiparticle
tunnelling~\cite{averin,nakamura,fitzgerald,choi,djqp1,djqp2}.

In this paper we analyse the back-action of a SSET on a
nanomechanical resonator. In particular, we investigate the
dynamics of a resonator coupled as a voltage-gate to a SSET which
is tuned in the vicinity of the JQP or the DJQP resonance. We find
that for both resonances the resonator can act as though it were
coupled to a thermal bath with its dynamics  described by a
Fokker-Planck equation, as was found for the normal state SET.
However, the magnitudes of the effective temperature and damping
of the resonator in the vicinity of the JQP and DJQP resonances
differ substantially, both from each other and from those for the
normal state case. A resonator coupled to a normal state SET has
an effective temperature which is proportional to the drain-source
voltage applied to the SET and is always damped. In contrast, the
effective temperature of a resonator coupled to a SSET in the
vicinity of the JQP or DJQP resonance is largely controlled by how
far the SSET is biased from the resonance, rather than the
magnitude of the drain-source voltage, and hence can easily be an
order of magnitude smaller than the effective temperature for an
analogous normal state SET. Furthermore, as the applied gate and
drain-source voltages are adjusted to tune the SSET through a
given JQP resonance, we find that the Fokker-Planck description
breaks down sufficiently close to the resonance, while further
from the resonance on the other side the Fokker-Planck description
is restored once again, but with negative effective temperature
and resonator damping constant implying the possibility of a
dynamic instability. Very similar results for the SSET-resonator
system have also been obtained independently using a different
approach by Clerk and Bennett~\cite{Clerk3}.

The dynamics of a nanomechanical resonator coupled to a SSET in the
vicinity of the JQP resonance bears a striking resemblance to a number of
other physical systems. In particular, the behaviour of the
effective temperature of the resonator in the vicinity of the JQP resonance
takes a very similar form to that of atoms undergoing Doppler
cooling due to their interactions with laser
light~\cite{stenholm,lett}. Indeed, the minimum effective
temperature of both a resonator in the vicinity of the JQP resonance and
Doppler-cooled atoms are given by a decay rate: the quasiparticle
decay rate for the SSET-resonator system and the decay rate of the
excited state for the atoms.

This paper is organized as follows. In section 2, we introduce a
master equation describing the coupled statistical dynamics of the
SSET-resonator system in the vicinity of the JQP resonance (a similar master
equation for the SSET-resonator system in the vicinity of the DJQP resonance
is described in the appendix). We then show that the master
equation can be well-approximated by a Fokker-Planck equation
under conditions of weak coupling and wide separation of SSET and
oscillator dynamics timescales. In section 3 we present analytic
and numerical calculations of the SSET-induced damping, frequency
renormalization and effective temperature in the vicinity of the JQP and
DJQP resonances. In section 4, we discuss our results and
the analogy between the SSET-resonator device and other physical
systems, before we finally present our conclusions.

\section{Master Equation description for the JQP resonance}

In this section we obtain a master equation for the
SSET-resonator system in the vicinity of the JQP resonance and show that the
dynamics of the resonator can be described by a Fokker-Planck
equation. The same approach can also be used to derive analogous
results for the DJQP resonance, details of which are given in
the appendix.

\begin{figure}[t]
\centering \epsfig{file=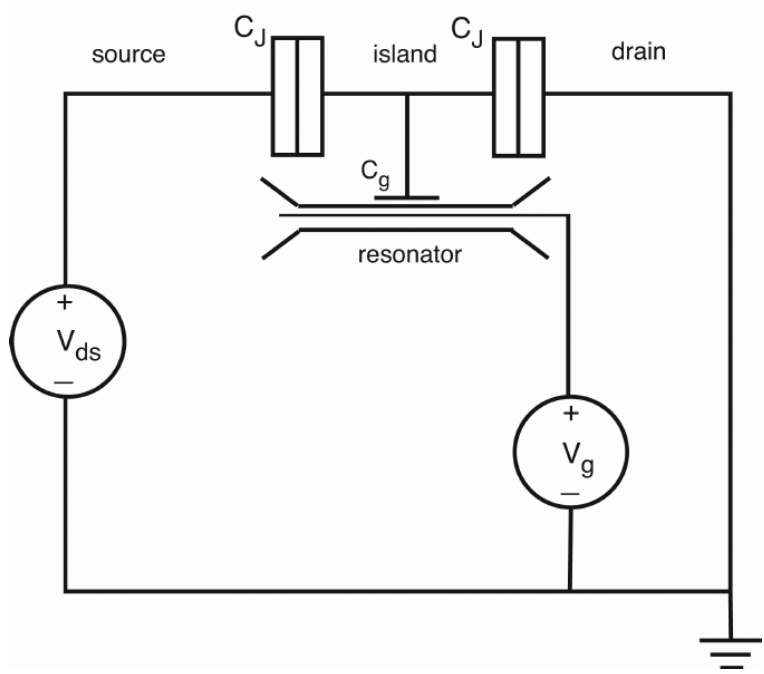,width=10cm} \caption{Model
circuit of the SSET-resonator system.} \label{model}
\end{figure}

The model circuit that we consider is shown in figure~\ref{model}. The SSET
consists of a small superconducting island and two superconducting
leads weakly-linked to the island via tunnel junctions with
capacitances $C_J$; a drain-source bias voltage $V_{ds}$ is
applied to the leads. The nanomechanical resonator is treated as a
single-mode harmonic oscillator with frequency $\omega$ and mass
$m$. The metallized resonator is located adjacent the SSET island,
forming a gate  with capacitance $C_g(x)$, which depends on the
resonator's displacement $x$; a gate voltage $V_{g}$ is applied to
the resonator. In the experiments of Refs.~\cite{mset1,mset2}, the
coupling between the SSET island and the resonator is typically
very weak so that the displacement of the resonator from its
equilibrium position is much less than the separation $d$ between
that equilibrium position and the SSET island itself. Hence if we
also assume a parallel-plate geometry for simplicity, the gate
capacitance can be approximated by~\cite{armour}:
$C_g(x)=C_g(1-x/d)$, implying linear coupling between the SSET and
the resonator~\cite{Clerk2}.

The central island of the SSET is taken to be sufficiently small
that its charging energy $E_{c}=e^{2}/2(2C_{J}+C_{g})\sim\Delta\gg
k_{\rm B}T$,
where $\Delta$ is the
superconducting gap and $T$ is the temperature of the quasiparticles in the
leads. Hence, the number of charge states accessible to the island
is severely restricted. The Josephson coupling between the leads and the
island is $E_J=h\Delta/(8e^2R_J)\ll E_c$, where  $R_J$ the resistance of the
junctions~\cite{djqp1}. Depending on the exact value of the
polarization charge induced on the SSET island by the resonator
gate, $N_{g}=(C_{g}V_{g}+C_{J}V_{ds})/e$, certain quasiparticle
and resonant Cooper pair tunnelling processes can become
energetically favourable leading to a number of possible current
carrying regimes such as the JQP and DJQP cycles. At sufficiently
large drain-source voltages, and for relatively low junction
resistances, it is also possible for current to flow via higher
order processes such as quasiparticle co-tunnelling, but we will
neglect such effects in what follows.

\begin{figure}[t]
\centering {
\epsfig{file=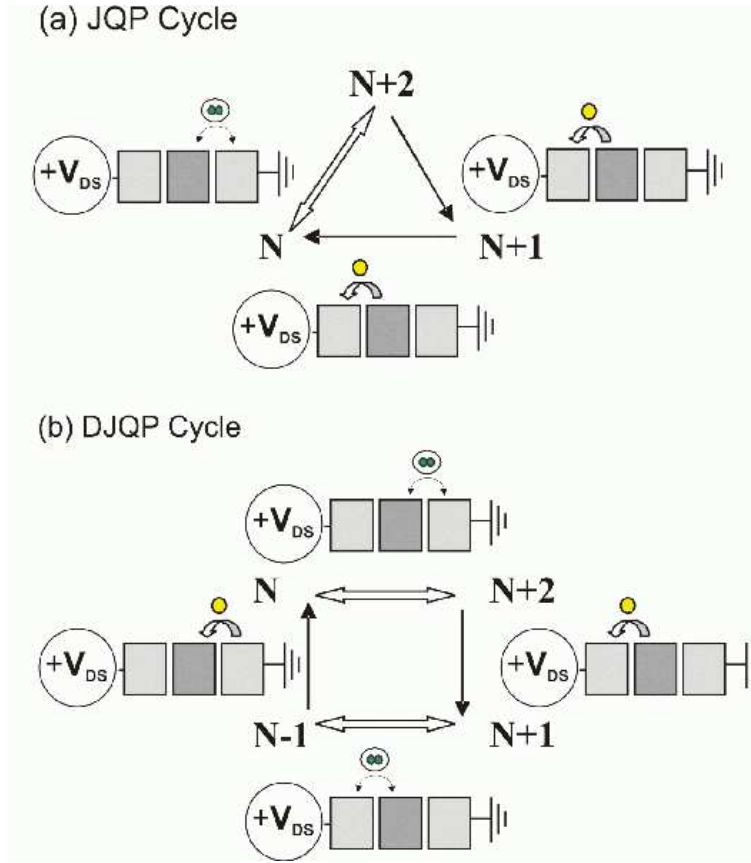,width=10cm}}\caption{Schematic
illustration of the JQP  and DJQP cycles.  (a) For the JQP cycle,
Josephson tunnelling involving a Cooper pair occurs between drain
and island electrodes, increasing the island electron number $N$
by 2, followed by two, subsequent quasiparticle tunnel decay
processes into the source electrode, decreasing the electron
island number by 2~\cite{choi}. (b) For the DJQP cycle, Josephson
tunnelling of a Cooper pair occurs at both junctions, with a
Cooper pair tunnelling event through a given junction alternating
in turn with a quasiparticle tunnelling event at the opposite
junction~\cite{djqp1}.} \label{JQP}
\end{figure}

The details of the specific electronic processes which occur at
the JQP and DJQP resonances are illustrated schematically in
figure~\ref{JQP}. Close to the JQP resonance, Cooper pairs tunnel between
the right (with reference to the circuit in Figure~\ref{model}), drain
electrode and island, while electron quasiparticles tunnel out
from the island to the left, source electrode. Alternatively,
Cooper pair tunnelling can occur between the left source
electrode and island, while electron quasiparticles tunnel in from
the right drain electrode to the island. Which of these two JQP cycles
takes place depends on the gate and drain-source voltage biases.
We shall consider biases such that only the former cycle occurs
(i.e., that illustrated in figure~\ref{JQP}a).

We seek a master equation that describes the dynamics of the
island charge state of the SSET and the position-velocity state of
the resonator's center-of-mass valid in the vicinity of the JQP
resonance. Master equations for the island charges in normal state
and superconducting SETs have been derived by a number of groups
using essentially the same technique~\cite{schoeller,choi}.
Starting with the full (time-dependent) Schr\"{o}dinger equation
for the system, one traces over the microscopic electronic degrees
of freedom making use of a second-order Born approximation (that
treats the quasiparticle tunnelling rates between the island and
the leads as a small expansion parameter), followed by a long-time
limit Markov approximation (that treats the response of the
electrons in the leads to a tunnelling event as being
instantaneous), to arrive at a master equation for a reduced
density matrix represented in the basis of total number of
electrons, $N$, on the island. We generalise this approach to
include the resonator. Again starting from a fully quantum
Hamiltonian, we follow the same procedure of the Born and Markov
approximations and trace over the microscopic degrees of freedom.
However, we also assume that the resonator does not evolve at all
on the time-scale of the quasiparticle tunnelling processes.
Essentially this means that we are treating the resonator as a
classical oscillator\footnote{In deriving the quasiparticle
tunnelling terms we  effectively treat the resonator as a
classical oscillator by making an adiabatic approximation, i.e.\
we assume that it does not move on the time-scale of the
tunnelling processes, an approximation which was also used in
Ref.~\cite{armour}. The other terms in the master equations which
arise from the coherent evolution of the resonator and SSET charge
are not affected by the adiabatic approximation. As we shall see
later on, the position dependence of the quasiparticle transition
rates in fact do not play an important role in affecting the
resonator dynamics near the JQP resonance so we expect our master
equations to provide a description of the resonator dynamics close
to the JQP resonance which is essentially the same as that which
would be obtained from a fully quantum derivation.}. Finally we
take the Wigner transform~\cite{agarwal} of the resulting
equations to obtain the desired master equation which can be
thought of as providing a semiclassical description of the coupled
dynamics~\cite{chemphys}.

For the JQP process (see figure~\ref{JQP}a), the island electron number can
be $N$, $N+1$ or $N+2$, with the $N$ and $N+2$ number states linked by coherent
Cooper pair tunnelling. Hence the associated set of coupled
semiclassical master equations has diagonal components
$\rho_{N}(x,v,t)$, $\rho_{N+1}(x,v,t)$, $\rho_{N+2}(x,v,t)$, and
off-diagonal component
$\rho_{N,N+2}(x,v,t)=\rho^{*}_{N+2,N}(x,v,t)$, where  $x$ and $v$
are the position and velocity coordinates, respectively, of the
oscillator. In our semiclassical description, the sum
$\rho_{N}(x,v,t)+\rho_{N+1}(x,v,t)+\rho_{N+2}(x,v,t)$ is the
probability density $P_{\rm HO}(x,v,t)$ of finding the oscillator at
the point in phase space $(x,v)$ at time $t$, while the integral
$\int dx dv \rho_{N}(x,v,t)$ gives the probability $P_{N}(t)$ that
the island electron number is $N$ at time $t$, with the
probability conservation $P_{N}+P_{N+1}+P_{N+2}=1$. Explicitly,
the semiclassical master equations take the form
\begin{eqnarray}
    \dot{\rho}_{N}&=&\omega^{2}(x+Nx_{s})\frac{\partial\rho_{N}}
    {\partial v}-v\frac{\partial\rho_{N}}{\partial x}
    +i\frac{E_{J}}{2\hbar}\left(\rho_{N+2,N}-\rho_{N,N+2}\right)\cr
    &+&\left[\Gamma(E_{N+1,N})+\Gamma'(E_{N+1,N})m\omega^{2}x_{s}x\right]
    \rho_{N+1}\cr
    \dot{\rho}_{N+2}&=&\omega^{2}[x+(N+2)x_{s}]\frac{\partial\rho_{N+2}}
    {\partial v}-v\frac{\partial\rho_{N+2}}{\partial x}
    -i\frac{E_{J}}{2\hbar}\left(\rho_{N+2,N}-\rho_{N,N+2}\right)\cr
    &-&\left[\Gamma(E_{N+2,N+1})+\Gamma'(E_{N+2,N+1})m\omega^{2}x_{s}x\right]
    \rho_{N+2}\cr
    \dot{\rho}_{N+1}&=&\omega^{2}[x+(N+1)x_{s}]\frac{\partial\rho_{N+1}}
    {\partial v}-v\frac{\partial\rho_{N+1}}{\partial x} \cr
    &+&\left[\Gamma(E_{N+2,N+1})+\Gamma'(E_{N+2,N+1})m\omega^{2}x_{s}x\right]
    \rho_{N+2}\cr
    &-&\left[\Gamma(E_{N+1,N})+\Gamma'(E_{N+1,N})m\omega^{2}x_{s}x\right]
    \rho_{N+1}\cr
    \dot{\rho}_{N,N+2}&=&\omega^{2}[x+(N+1)x_{s}]\frac{\partial\rho_{N,N+2}}
    {\partial v}-v\frac{\partial\rho_{N,N+2}}{\partial x}
    +i\frac{E_{J}}{2\hbar}\left(\rho_{N+2}-\rho_{N}\right)\cr
    &+&\frac{i}{\hbar}\left(E_{N+2,N}+2m\omega^{2}x_{s}x\right)\rho_{N,N+2}\cr
    &-&\frac{1}{2}
    \left[\Gamma(E_{N+2,N+1})+\Gamma'(E_{N+2,N+1})m\omega^{2}x_{s}x\right]
    \rho_{N,N+2},
    \label{mastereq}
\end{eqnarray}
where $x_{s}=2E_{c}C_{g}V_g/(em\omega^{2}d)$, which  is the
distance between equilibrium positions of the oscillator with $N$
and $N+1$ electrons on the island, parametrises the strength of
the electro-mechanical coupling. The quasiparticle tunnel rates
are given by~\cite{nak2}
\begin{equation}
\Gamma(E)=\frac{1}{e^{2} R_{J}}\int_{-\infty}^{\infty}d\epsilon
\rho(\epsilon) \rho(\epsilon+E) n(\epsilon)[1-n(\epsilon+E)],
\label{tunnelrate}
\end{equation}
where
\begin{equation}
    \rho(\epsilon)=\sqrt{\frac{\epsilon^{2}}{\epsilon^{2}-\Delta^{2}}}\Theta
(\epsilon^{2}-\Delta^{2}) \label{dos}
\end{equation}
is the normalized quasiparticle density of states, with
$\Theta(\cdot)$ the stepfunction, and
$n(\epsilon)=1/\left[1+\exp(\epsilon/k_{\rm B}T)\right]$. In the above
master equations, the typical position coordinate is assumed
sufficiently small that the tunnel rates can be expanded to first
order in the position coordinate, with $\Gamma'(E)=d\Gamma/dE$.
The quantities $E_{N+2,N+1}$ and $E_{N+1,N}$ are the energies
gained by an electron when it tunnels from island to source
electrode, the island number changing from $N+2$ to $N+1$ and
$N+1$ to $N$, respectively. The quantity $E_{N+2,N}$ is the energy
gained by a Cooper pair when it tunnels from island to drain
electrode. These energies are as follows:
\begin{eqnarray}
    E_{N+2,N+1}&=&-2E_{c}(N_{g}-N-3/2)+eV_{ds}\cr
    E_{N+1,N}&=&-2E_{c}(N_{g}-N-1/2)+eV_{ds}\cr
    E_{N+2,N}&=&-4E_{c}(N_{g}-N-1).\label{energies}
\end{eqnarray}
The JQP resonance condition is $E_{N+2,N}=0$ which is satisfied
for $N_{g}=N+1$. Furthermore, the bias voltage must be
sufficiently large for the quasiparticle processes to be allowed,
enabling the JQP cycle. For superconductors at zero temperature,
this translates into the requirement that $eV_{ds}> 2\Delta +E_c$.
Notice also that in the absence of the mechanical oscillator, we
recover the standard master equations for the SSET about the JQP
resonance~\cite{choi}.

It is convenient to express the master equation in terms of
dimensionless coordinates, since in dimensionless form the
essential parameters governing the dynamics are more clearly
expressed. Rewriting the time coordinate in units of the
tunnelling time, $\tau_{\rm tunnel}=eR_{J}/V_{ds}$, the position
coordinate in units of $x_{s}$, and the velocity coordinate in
units of $x_{s}/\tau_{\rm tunnel}$, the master equations take the
form
\begin{eqnarray}
    \dot{\rho}_{N}&=&\epsilon_{\rm HO}^{2}(x+N)\frac{\partial\rho_{N}}
    {\partial v}-v\frac{\partial\rho_{N}}{\partial x}
    +i\pi\epsilon_{J}\left(\rho_{N+2,N}-\rho_{N,N+2}\right)\cr
    &&+\left[\tilde{\Gamma}(\tilde{E}_{N+1,N})
    +\tilde{\Gamma}'(\tilde{E}_{N+1,N})\kappa x\right]
    \rho_{N+1}\cr
    \dot{\rho}_{N+2}&=&\epsilon_{\rm HO}^{2}[x+(N+2)]\frac{\partial\rho_{N+2}}
    {\partial v}-v\frac{\partial\rho_{N+2}}{\partial x}
    -i\pi\epsilon_{J}\left(\rho_{N+2,N}-\rho_{N,N+2}\right)\cr
    &&-\left[\tilde{\Gamma}(\tilde{E}_{N+2,N+1})+
    \tilde{\Gamma}'(\tilde{E}_{N+2,N+1})\kappa x\right]
    \rho_{N+2}\cr
    \dot{\rho}_{N+1}&=&\epsilon_{\rm HO}^{2}[x+(N+1)]\frac{\partial\rho_{N+1}}
    {\partial v}-v\frac{\partial\rho_{N+1}}{\partial x}\cr
    &&+\left[\tilde{\Gamma}(\tilde{E}_{N+2,N+1})+
    \tilde{\Gamma}'(\tilde{E}_{N+2,N+1})\kappa x\right]
    \rho_{N+2}\cr
    &&-\left[\tilde{\Gamma}(\tilde{E}_{N+1,N})+
    \tilde{\Gamma}'(\tilde{E}_{N+1,N})\kappa x\right]
    \rho_{N+1}\cr
    \dot{\rho}_{N,N+2}&=&\epsilon_{\rm HO}^{2}[x+(N+1)]
    \frac{\partial\rho_{N,N+2}}
    {\partial v}-v\frac{\partial\rho_{N,N+2}}{\partial x}
    +i\pi\epsilon_{J}\left(\rho_{N+2}-\rho_{N}\right)\cr
    &&+2\pi i r\left(\tilde{E}_{N+2,N}+2\kappa x\right)\rho_{N,N+2}\cr
    &&-\frac{1}{2}\left[\tilde{\Gamma}(\tilde{E}_{N+2,N+1})+
    \tilde{\Gamma}'(\tilde{E}_{N+2,N+1})\kappa x\right]
    \rho_{N,N+2}.\label{mastereq2}
\end{eqnarray}
The definitions of the various dimensionless parameters are as
follows: $\epsilon_{\rm HO}=\omega \tau_{\rm tunnel}$ is the ratio of
the SET quasiparticle tunnelling time to the oscillator period,
$\epsilon_{J}=\tau_{\rm tunnel}/\tau_{\rm Rabi}=(e R_{J}/V_{ds})
(E_{J}/h)=\Delta/(8eV_{ds})$ is the ratio of the quasiparticle
tunnelling time to the Cooper pair Rabi oscillation period,
$r=R_{J}/(h/e^{2})$ is the ratio of the tunnel junction resistance
to the quantum of resistance, and $\kappa=m\omega^{2}
x_{s}^{2}/(eV_{ds})$ characterizes the coupling strength between
the oscillator and the SET. The dimensionless tunnelling rate is
\begin{equation}
    \tilde{\Gamma}(\tilde{E})=\int_{-\infty}^{\infty}d\tilde{\epsilon}
\rho({\tilde{\epsilon}}) \rho({\tilde{\epsilon}}+\tilde{E})
n({\tilde{\epsilon}})[1-n({\tilde{\epsilon}}+\tilde{E})],\label{tunnelrate2}
\end{equation}
where now
\begin{equation}
    \rho({\tilde{\epsilon}})=\sqrt{\frac{{\tilde{\epsilon}}^{2}}
{{\tilde{\epsilon}}^{2}-\tilde{\Delta}^{2}}}\Theta
({\tilde{\epsilon}}^{2}-\tilde{\Delta}^{2})\label{dos2}
\end{equation}
and $n(\tilde{\epsilon})=1/[1+\exp(\tilde{\epsilon} e
V_{ds}/k_{\rm B}T)]$, with $\tilde{\epsilon}=\epsilon/(eV_{ds})$,
$\tilde{E}=E/(eV_{ds})$, and $\tilde{\Delta}=\Delta/(eV_{ds})$.
Note that for typical nanomechanical-SSETs~\cite{mset1,mset2}, we
have $\epsilon_{\rm HO}\ll\epsilon_{J}\ll 1$, $\kappa\ll 1$, and $r\ge
1$. Also, the dimensionless quasiparticle tunnel rates
$\tilde{\Gamma}$, and their gradients $\tilde{\Gamma}'$, are
generally of order unity.

Our goal is to obtain a description of the dynamics of the
resonator, decoupled from the details of the electronic degrees of
freedom. One very direct way to obtain the resonator dynamics is
to solve numerically the above master equation for the oscillator
probability density $P_{\rm HO}(x,v,t)$. Another direction is to take
advantage of the typical conditions of weak coupling ($\kappa\ll
1$) and wide separation of timescales ($\epsilon_{\rm HO}\ll 1$) to
derive from the above master equation a much simpler, approximate
effective equation for the oscillator probability density
$P_{\rm HO}(x,v,t)$ (i.e.\ a reduced master equation for the oscillator
alone) which can then be easily solved. We have used both
approaches. In the remainder of this section we describe how the
reduced master equation is obtained and show that it is nothing
other than the Fokker-Planck equation. Later in section 3 we compare the
results of this approach with direct numerical integrations of the
original set of master equations.

We begin our derivation of the reduced master equation of the
resonator by rewriting the full set of the master equations
(\ref{mastereq2}) in the following, concise $5\times 5$ matrix
operator form:
\begin{equation}
    \dot{\cal P}=\left({\cal H}_{0}+{\cal V}\right){\cal P},
    \label{schrodingereq}
\end{equation}
where
\[{\cal P}=\left(\begin{array}{c} \rho_{N+2}(x,v,t)\\ \rho_{N}(x,v,t)\\
\rho_{N+1}(x,v,t)\\ {\rm Im}~\rho_{N,N+2}(x,v,t)\\ {\rm Re}~\rho_{N,N+2}(x,v,t)
\end{array}\right),
\]
\begin{eqnarray}
    {\cal H}_{0}&=&
    \left(\epsilon_{\rm HO}^{2}x\frac{\partial}{\partial
    v}-v\frac{\partial}{\partial x}\right){\cal I} \\
    &+&\left(\begin{array}{ccccc}
    -\Gamma_{N+2,N+1} & 0 & 0 & -2\pi\epsilon_{J} &0\\
     0 & 0 & \Gamma_{N+1,N} & 2\pi\epsilon_{J} & 0\\
     \Gamma_{N+2,N+1} & 0 & -\Gamma_{N+1,N} & 0 & 0\\
     \pi\epsilon_{J} & -\pi\epsilon_{J} & 0 & -\frac{\Gamma_{N+2,N+1}}{2} &
     2\pi r E_{N+2,N}\\
     0 & 0 & 0 & -2\pi r E_{N+2,N} & -\frac{\Gamma_{N+2,N+1}}{2}
    \end{array}\right)\nonumber
    \label{H0}
\end{eqnarray}
with ${\cal{I}}$ denoting the $5\times 5$ identity matrix and we
use the shorthand notation ${\Gamma}(\tilde{E}_{N+2,N+1})\equiv
\Gamma_{N+2,N+1}$ (we have dropped the tilde for convenience,
understanding that all quantities are in dimensionless form). The
operator describing the interaction between the SET and oscillator
is ${\cal V}={\cal V}_{1}+{\cal V}_{2}=\kappa x{\cal U}_{1}+
\epsilon_{\rm HO}^{2}\frac{\partial}{\partial v}{\cal U}_{2}$ where
\begin{equation}
    {\cal U}_{1}=
    \left(\begin{array}{ccccc}
    -\Gamma'_{N+2,N+1} & 0 & 0 & 0 & 0\\
    0 & 0 & \Gamma'_{N+1,N} & 0 & 0\\
    \Gamma'_{N+2,N+1} & 0 & -\Gamma'_{N+1,N} & 0 & 0\\
    0 & 0 & 0 & - \frac{\Gamma'_{N+2,N+1}}{2} & 4\pi r\\
    0 & 0 & 0 & -4\pi r & -\frac{\Gamma'_{N+2,N+1}}{2}
    \end{array}\right)
    \label{U1}
\end{equation}
and
\begin{equation}
    {\cal U}_{2}=
   \left(\begin{array}{ccccc}
    1+\Delta P & 0 & 0 & 0 & 0\\
    0 & -1+\Delta P & 0 & 0 & 0\\
    0 & 0 & \Delta P& 0 & 0\\
    0 & 0 & 0 & \Delta P& 0\\
    0 & 0 & 0 & 0 & \Delta P
    \end{array}\right),
    \label{U2}
\end{equation}
where we have used the shorthand notation $\Delta P=\langle
P_{N}\rangle-\langle P_{N+2}\rangle$. Note that we have redefined
the position coordinate such that its origin coincides with the
steady-state value $\langle x\rangle = -(N+1) +\langle
P_{N}\rangle-\langle P_{N+2}\rangle$, where the steady state
island occupation probabilities are taken to be those for the
limit $\kappa\rightarrow 0$,
\begin{eqnarray}
    \langle P_{N+2}\rangle
    &=&\frac{(\pi\epsilon_{J})^{2}}{(\Gamma_{N+2,N+1}/2)^{2}+(2\pi r
    E_{N+2,N})^{2}+(\pi\epsilon_{J})^{2}
    \left(2+\frac{\Gamma_{N+2,N+1}}{\Gamma_{N+1,N}}\right)}\cr
    \langle P_{N+1}\rangle
    &=&\frac{\Gamma_{N+2,N+1}}{\Gamma_{N+1,N}}\langle
    P_{N+2}\rangle \nonumber\\
    \langle P_{N}\rangle &=& 1-\langle P_{N+1}\rangle-\langle
    P_{N+2}\rangle,\label{steadystprob}
\end{eqnarray}
an approach which is valid for sufficiently weak coupling. The
advantage of working with this redefined position coordinate will
become apparent shortly.

Equation~(\ref{schrodingereq}) resembles the time-dependent
Schr\"{o}dinger equation. The `Hamiltonian operator' ${\cal
H}_{0}$ gives the free, decoupled evolution of the independent
oscillator and SSET systems, while the operator ${\cal V}={\cal
V}_{1}+{\cal V}_{2}$ describes the interaction between the two
systems with ${\cal V}_{1}$ giving the dependence of the
Cooper-pair and quasiparticle tunnelling rates on the oscillator
position and ${\cal V}_{2}$ giving the SSET island number
dependence of the electrostatic force acting on the oscillator.

Given the close resemblance of equation~(\ref{schrodingereq})
to the Schr\"{o}dinger equation, we can apply approximation techniques developed
for open quantum systems, in particular the
self-consistent Born approximation (SCBA), followed by the Markov
approximation. Applying the SCBA as
described in section 3.1 of Ref.~\cite{paz1}, assuming weak coupling
between the oscillator and SSET, $\kappa\ll
1$, we obtain the following approximate expression for the reduced
master equation probability distribution $P_{\rm HO}(x,v,t)$ of
the oscillator~\cite{blencowe}:
\begin{eqnarray}
\fl \dot{P}_{\rm HO}(x,v,t)={\cal H}_{\rm HO} P_{\rm HO}(x,v,t)
    +e^{{\cal H}_{\rm HO}t}{\rm Tr}_{\rm SET}\left[{\cal V}(t) {\cal P}_{\rm
    SET}(0)\right]e^{-{\cal H}_{\rm HO}t}P_{\rm HO}(x,v,t)\cr
     \fl -\int_{0}^{t}dt' ~e^{{\cal H}_{\rm HO}t}
    {\rm Tr}_{\rm SET}\left[{\cal V}(t) {\cal P}_{\rm
    SET}(0)\right]{\rm Tr}_{\rm SET}\left[{\cal V}(t') {\cal P}_{\rm
    SET}(0)\right]e^{-{\cal H}_{\rm HO}t}P_{\rm HO}(x,v,t)\cr
   \fl +\int_{0}^{t}dt' ~e^{{\cal H}_{\rm HO}t}
    {\rm Tr}_{\rm SET}\left[{\cal V}(t){\cal V}(t') {\cal P}_{\rm
    SET}(0)\right]e^{-{\cal H}_{\rm HO}t}P_{\rm HO}(x,v,t),
    \label{SCBA}
\end{eqnarray}
where ${\cal H}_{\rm HO}=\epsilon^{2}x\frac{\partial}{\partial v}-
v\frac{\partial}{\partial x}$ is the Hamiltonian operator for the
free harmonic oscillator and ${\cal V}(t)=e^{-{\cal H}_{0}t}{\cal
V}e^{+{\cal H}_{0}t}$ is in the interaction picture. The initial,
$t=0$  probability distribution is taken to be a product state:
${\cal P}(0)= P_{\rm HO}(x,v,0){\cal P}_{\rm SET}(0)$, where
\[{\cal
P}_{\rm SET}(0)=\left(\begin{array}{c}
P_{N+2}(0)\\ P_{N}(0)\\ P_{N+1}(0)\\{\rm Im}~P_{N,N+2}(0)\\
{\rm Re}~P_{N,N+2}(0)
\end{array}\right).\]
Note that the above SCBA step which gives the oscillator master
equation~(\ref{SCBA})  and also the Markov approximation applied below
should not be
confused with the  Born-Markov approximation  described earlier in section 2
which gives the starting oscillator-SSET master equation~(\ref{mastereq});
these two approximation steps rely on distinct weak coupling and timescale
conditions.

The SCBA approach was applied to obtain a Fokker-Planck equation
for a resonator coupled to a normal state SET in Ref.~\cite{blencowe} and our
derivation for the SSET case follows the
same route. We can use a Markov approximation and evaluate the
integrals in~(\ref{SCBA}) for $t\rightarrow\infty$ since typically
$\epsilon_{\rm HO}\ll 1$ and we are only interested in the oscillator
dynamics on timescales of order the mechanical period and longer,
$t\gtrsim\epsilon_{\rm HO}^{-1}$. Furthermore, using the redefined position
coordinate, we find that the second and third terms on the right
hand side of equation~(\ref{SCBA}) drop out and we eventually
obtain
\begin{eqnarray}
    \frac{\partial P_{\rm HO}}{\partial t}&=&
    \left[\epsilon_{\rm HO}^{2}x\frac{\partial}{\partial v}-
    v\frac{\partial}{\partial x} +\kappa\epsilon_{\rm HO}^{2}
    \frac{\partial}{\partial v}\left({\cal C}_{1}^{21}x-{\cal
    C}_{2}^{21}v\right)\right.\cr
    &&\left.+\epsilon_{\rm HO}^{4}\frac{\partial}{\partial v}
    \left({\cal C}_{1}^{22}\frac{\partial}{\partial
    v}+{\cal C}_{2}^{22}\frac{\partial}{\partial  x}\right)\right] P_{\rm HO},
    \label{fokkerplanck}
\end{eqnarray}
where the ${\cal C}$ coefficients are defined as follows:
\begin{equation}
    {\cal C}_{1}^{ij}=\lim_{t\rightarrow\infty}\int_{0}^{t}
    d\tau {\rm Tr}\left[{\cal
    U}_{i}(t){\cal U}_{j}(t-\tau) {\cal P}_{\rm
    SET}(0)\right]
    \label{C1}
\end{equation}
and
\begin{equation}
    {\cal C}_{2}^{ij}=\lim_{t\rightarrow\infty}\int_{0}^{t}
    d\tau \tau{\rm Tr}\left[{\cal
    U}_{i}(t){\cal U}_{j}(t-\tau) {\cal P}_{\rm
    SET}(0)\right].
    \label{C2}
\end{equation}
In the Markovian limit which we have used, these coefficients do
not depend on the initial state ${\cal P}_{\rm SET}(0)$  of the
SET, just as we would expect. Furthermore, the $\partial^{2}
P/\partial v\partial x$ term in equation~(\ref{fokkerplanck})
(called the `anomalous diffusion' term in Ref.~\cite{paz1}) is of
order $\epsilon$ smaller than the diffusion term when time is
expressed in units of the oscillator period; it should have only a
small effect on timescales of order the mechanical period or
longer. Re-expressing equation~(\ref{fokkerplanck}) in terms of
dimensionful coordinates and dropping the anomalous diffusion
term, we obtain
\begin{equation}
     \frac{\partial P_{\rm HO}}{\partial
     t}=\left[\omega_{R}^{2}x\frac{\partial}{\partial v}
     -v\frac{\partial}{\partial x}+\gamma_{\rm SET}\frac{\partial}{\partial v}v
     +\frac{\gamma_{\rm SET} k_{\rm B}T_{\rm SET}}{m}
     \frac{\partial^{2}}{\partial v^{2}}\right] P_{\rm HO},
     \label{fokkerplanck2}
\end{equation}
where the renormalized oscillator frequency is
\begin{equation}
   \omega_{R}=\sqrt{1+\kappa {\cal C}_{1}^{21}}~\omega,
    \label{renormfreq}
\end{equation}
the damping rate is
\begin{equation}
    \gamma_{\rm SET}=-\kappa\epsilon\omega {\cal C}_{2}^{21},
    \label{dampingrate}
\end{equation}
and the effective SET temperature is
\begin{equation}
    k_{\rm B}T_{\rm SET}=-eV_{ds} \frac{{\cal C}_{1}^{22}}{{\cal C}_{2}^{21}}.
    \label{SETtemp}
\end{equation}

Equation~(\ref{fokkerplanck2}), which has the form of a particular
class of Fokker-Planck equation known as the Klein-Kramers
equation~\cite{risken}, describes the Brownian motion of a
harmonic oscillator interacting with a thermal bath. The
oscillator experiences a net damping force, due to the interaction
with the SSET, [the third term on the right-hand-side of
equation~(\ref{fokkerplanck2})] and an accompanying Gaussian
distributed thermal fluctuating force [the fourth term on the
right-hand-side of (\ref{fokkerplanck2}), called the `diffusion'
term].

Note that our derivation of the Fokker-Planck
equation~(\ref{fokkerplanck2}) for the oscillator does not in fact
depend on the specifics of the SET interacting with it. As long as
the original master equation has the form $\dot{\cal
P}=\left({\cal H}_{0}+{\cal V}\right){\cal P}$ with ${\cal
V}={\cal V}_{1}+{\cal V}_{2}=\kappa x{\cal U}_{1}+ \epsilon_{\rm
HO}^{2}\frac{\partial}{\partial v}{\cal U}_{2}$ where operators
${\cal U}_{1 (2)}$ involve only the SET parameters, then the same
effective thermal bath description~(\ref{fokkerplanck}) of the SET
results. The specifics of the SET enter in the dimensionless
${\cal C}$ coefficients defined in equations (\ref{C1}) and
(\ref{C2}). Thus, for the example of a normal state SET with
island electron number fluctuating between the values $N$ and
$N+1$, the coefficients take the values ${\cal C}_{1}^{21}={\cal
C}_{2}^{21}=-1$ and ${\cal C}_{1}^{22}=\langle P_{N}\rangle
\langle P_{N+1}\rangle$~\cite{armour,blencowe}. Furthermore, the
same approach can be applied to extract the relevant resonator
dynamics for the DJQP cycle (see section 3.3 and the appendix).

\section{Results}

The coefficients $\mathcal{C}_1^{ij}$ and $\mathcal{C}_2^{ij}$,
given by equations (\ref{C1}) and (\ref{C2}), together with the
Fokker-Planck equation (\ref{fokkerplanck}), describe the dynamics of a
resonator coupled to a SSET. In this section we evaluate equations
(\ref{C1}) and (\ref{C2}) numerically to obtain the effective
temperature, renormalized frequency and damping of the resonator
close to the JQP and DJQP resonances. In order to get a better
picture of the underlying physics, and also to understand the
limitations of the SCBA approach, we compare these numerical
results for the JQP approach with analytical approximations and
numerical results obtained by direct integration of the master
equations [given by equation (\ref{mastereq2})].

We begin this section by presenting approximate analytical results
derived for the JQP resonance as they provide a
useful framework within which to understand the resonator
dynamics. We then present our numerical results for the JQP and
DJQP resonances.

\subsection{Analytical approximations for the JQP resonance}

It is possible to derive analytical approximations to the ${\cal
C}$ coefficients as series expansions in $\epsilon_{J}$.
Neglecting also the $\Gamma'$ terms in the interaction operator
${\cal U}_{1}$, since they are an order of magnitude smaller than
the $4\pi r$ terms in ${\cal U}_{1}$, we find to leading
non-vanishing order in $\epsilon_{J}$:
\begin{eqnarray}
    &{\cal
    C}_{1}^{21}&=-\frac{(4\pi^{2}r\epsilon_{J})^{2}E_{N+2,N}
    \left(\Gamma_{N_+2,N+1}
    +2\Gamma_{N+1,N}\right)}
    {\Gamma_{N+1,N}\left[(\Gamma_{N+2,N+1}/2)^{2}+(2\pi r
    E_{N+2,N})^{2}\right]^{2}},
    \label{approxC121}
\end{eqnarray}
\begin{eqnarray}
   {\cal C}_{2}^{21}&=&
    -\frac{(4\pi^{2}r\epsilon_{J})^{2}E_{N+2,N}\Gamma_{N+2,N+1}}
    {\Gamma_{N+1,N}^{2}\left[(\Gamma_{N+2,N+1}/2)^{2}+(2\pi r
    E_{N+2,N})^{2}\right]^{3}}  \label{approxC221}\\
    &\times &\left[2\Gamma_{N+1,N}^{2}+(\Gamma_{N+2,N+1}/2)^{2}+
    \Gamma_{N+1,N}\Gamma_{N+2,N+1}+(2\pi r E_{N+2,N})^{2}\right],
\nonumber
\end{eqnarray}
\begin{eqnarray}
    {\cal C}_{1}^{22}&=&\frac{(\pi\epsilon_{J})^{2}\Gamma_{N+2,N+1}}
    {\Gamma_{N+1,N}^{2}\left[(\Gamma_{N+2,N+1}/2)^{2}+(2\pi r
    E_{N+2,N})^{2}\right]^{2}}   \label{approxC122}\\
    &\times &\left[2\Gamma_{N+1,N}^{2}+(\Gamma_{N+2,N+1}/2)^{2}+
    \Gamma_{N+1,N}\Gamma_{N+2,N+1}+(2\pi r E_{N+2,N})^{2}\right].
    \nonumber
\end{eqnarray}
The approximations to the renormalized frequency, damping rate and
effective SET temperature then follow by substituting
(\ref{approxC121}), (\ref{approxC221}), and (\ref{approxC122})
into equations (\ref{renormfreq}), (\ref{dampingrate}), and
(\ref{SETtemp}). In particular, for the SET temperature, we have:
\begin{equation}
    k_{\rm B}T_{\rm SET}=-e V_{ds}\frac{{\cal C}_{1}^{22}}{{\cal C}_{2}^{21}}=
   \frac{\hbar}{4} \frac{\Gamma_{N+2,N+1}^{2}+4(E_{N+2,N}/\hbar)^{2}}{4
    (E_{N+2,N}/\hbar)},
    \label{approxTSET}
\end{equation}
where $\Gamma_{N+2,N+1}$ and $E_{N+2,N}$ are in their original,
dimensional form [see equations~(\ref{tunnelrate}) and
(\ref{energies})]. Note that $T_{\rm SET}$ does not depend on
$\epsilon_{J}$ to leading order, while $\gamma_{\rm SET}$ and the
frequency renormalization are $O(\epsilon_{J}^{2})$.

Although these results are simple and intuitive, they are in
principle only valid when  $\pi \epsilon_J\ll 1$ a condition
which is by no means always satisfied for SSETs in practice.

\subsection{Numerical results for the JQP resonance}

Two distinct sets of numerical calculations are carried out. We
carry out integrations of the original master equations and within
the framework of the Born-Markov approximation (\ref{mastereq2}),
we evaluate the
expressions for the coefficients [equations (\ref{C1}) and
(\ref{C2})] numerically.

For the numerical integration of the master
equations~\cite{armour}, we obtain the full evolution of the resonator
probability distribution (with initial state chosen to be
Gaussian) from which we determine the evolution of the average
position (with respect to the fixed point value), $\langle
x(t)\rangle$. We then obtain values for $\omega_R$ and
$\gamma_{\rm SET}$ by fitting $\langle x(t)\rangle$ to the
equation of motion of a damped harmonic oscillator with a
renormalized frequency. Finally, we use  equations
(\ref{renormfreq}) and (\ref{dampingrate}) to infer values of the
corresponding coefficients. In obtaining these results we
concentrate  on the typical experimentally accessible regime
$\kappa,\epsilon \ll 1$. As a consequence, the numerical
integrations take a prohibitively long time to reach a
steady-state and so we do not extract values of $T_{\rm SET}$ from
the integrations.

\begin{figure} \centering {\epsfig{file=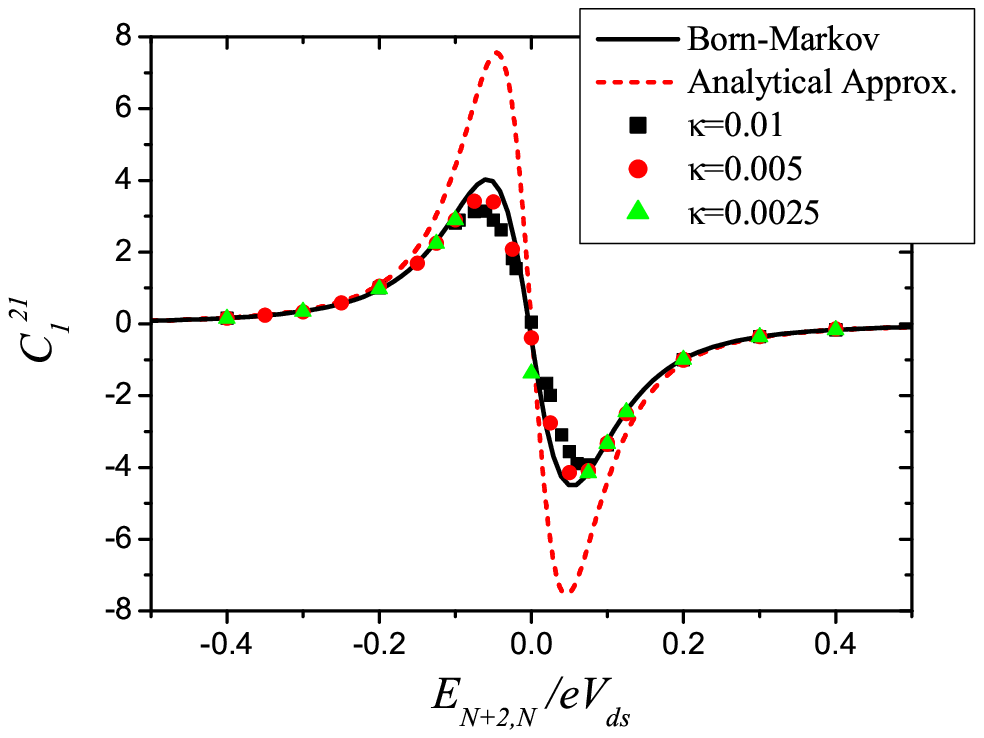,width=9cm}
\epsfig{file=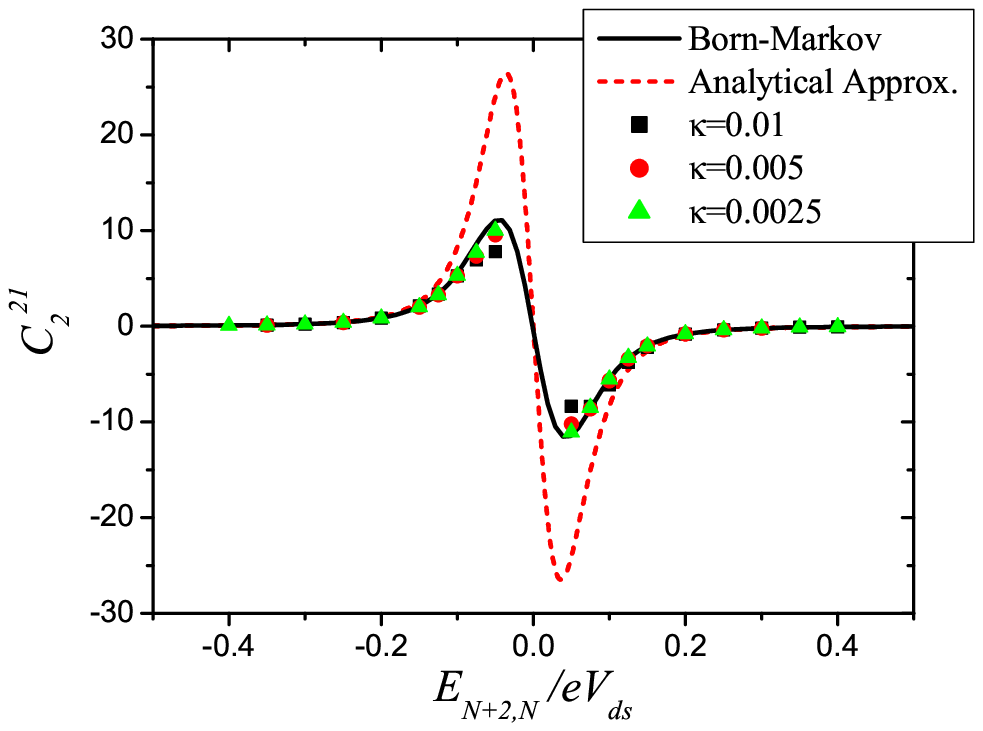,width=9cm}} \caption{Comparison of numerical
calculations of the coefficients $\mathcal{C}_1^{21}$ and
$\mathcal{C}_2^{21}$ with the analytical approximations [equations
(\ref{approxC121}) and (\ref{approxC221})]. The full curves come
from numerical evaluations of equations (\ref{C1}) and (\ref{C2}),
whereas the points come from numerical integrations of the master
equations (\ref{mastereq2}). For the numerics we choose
$\epsilon_J=1/16$, $r=1$ and set the quasiparticle tunnel rates
and their gradients to unity. In addition, when integrating the
master equations we set $\epsilon=0.1$ and vary the value of
$\kappa$ as shown.} \label{jqpc}
\end{figure}

\begin{figure}
\centering {\epsfig{file=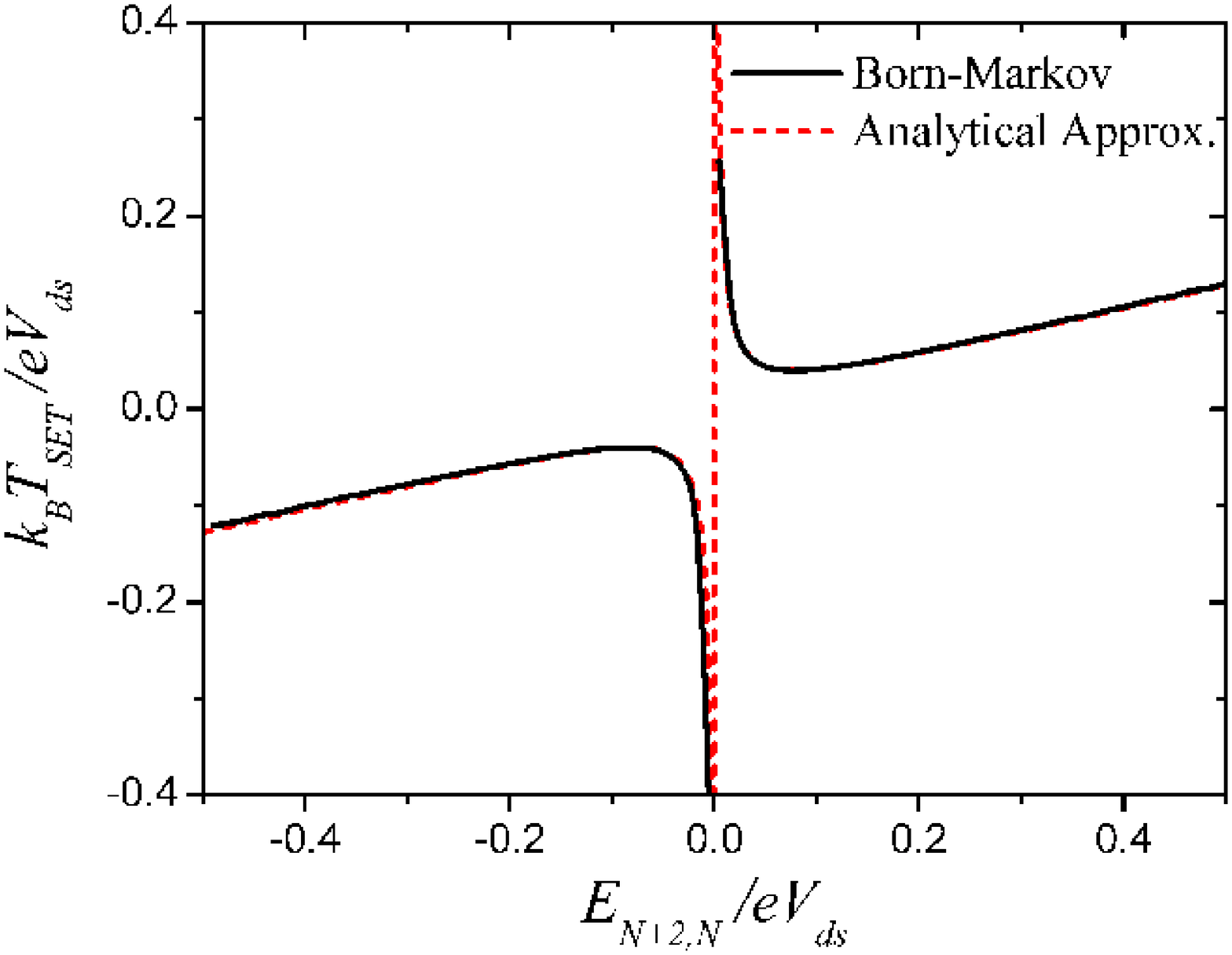,width=9cm}}
\caption{Comparison of numerical calculations of $T_{\rm SET}$
with the analytical approximation, equation (\ref{approxTSET}).
The parameters are the same as those used for figure~\ref{jqpc}.}
\label{tset}
\end{figure}

Figures~\ref{jqpc} and~\ref{tset} show our numerical calculations
of the coefficients $\mathcal{C}_1^{21}$, $\mathcal{C}_2^{21}$ and
$T_{\rm SET}$, with the corresponding analytical
approximations~(\ref{approxC121}), (\ref{approxC221})
and~(\ref{approxTSET}) shown for comparison. Notice that while the
numerical results for $\mathcal{C}_1^{21}$ and
$\mathcal{C}_2^{21}$ were obtained from  numerical
integrations of both the master equations (without Born-Markov
approximation) and from equations (\ref{C1}) and (\ref{C2}) (with
the Born-Markov approximation), the numerical results for $T_{\rm
SET}$ were only obtained using the latter technique. In the
numerics, values for the system parameters were chosen that are
typical of those found in current devices~\cite{mset1,mset2}. In
particular, the value of $\epsilon_J=\Delta/(8 e V_{ds})=1/16$ is
kept relatively small. This value corresponds to choosing $e
V_{ds}=2\Delta$, the correct order of magnitude to enable the JQP and
DJQP cycles. Nevertheless, it is clear from
figure~\ref{jqpc} that our analytical approximations differ
substantially from the numerics close to the centre of the
resonance. In contrast, figure~\ref{tset} shows that there is
excellent agreement between the analytical and numerical
calculations of $T_{\rm SET}$. It is not clear why the agreement
is so good, but the most obvious explanation is that the
cancellation of the $\epsilon_J$ dependent terms that occurs in
our approximate expression for the ratio
$\mathcal{C}_1^{22}/\mathcal{C}_2^{21}$ must extend beyond second
order in $\epsilon_J$.

\begin{figure}
\centering {\epsfig{file=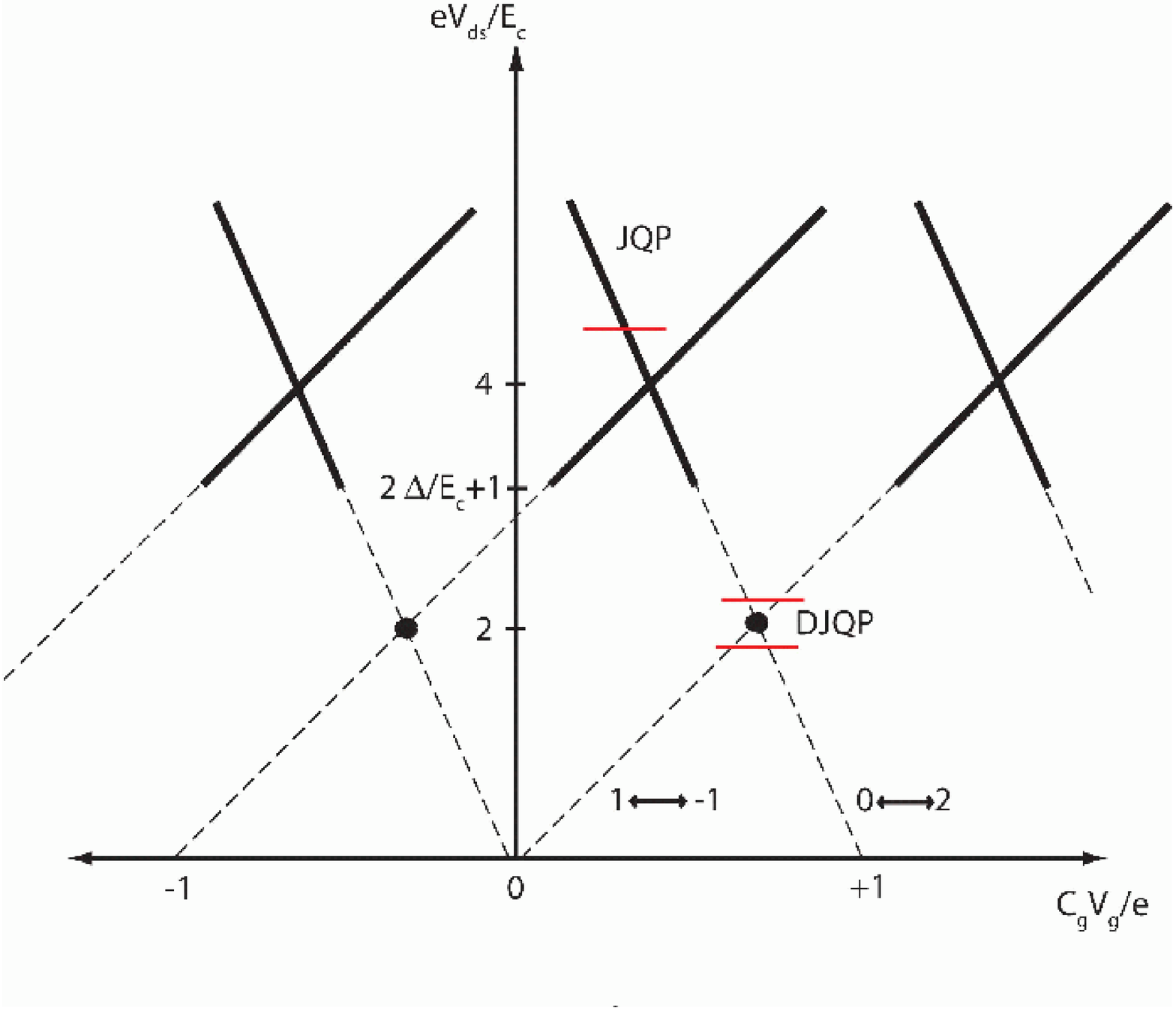,width=12cm}} \caption{Schematic
map showing the location of a selection of neighbouring JQP
resonances (solid black lines) and DJQP resonances (solid black
circles). The bias ranges for the $\mathcal{C}_1^{21}$,
$\mathcal{C}_2^{21}$ and $T_{\rm SET}$ plots are indicated by the
solid red lines.} \label{map}
\end{figure}

Figure~\ref{map} shows schematically the $V_{ds}$ and $V_{g}$ bias
ranges in relation to the JQP resonance lines for which the curves in
figures~\ref{jqpc} and~\ref{tset} are obtained. Note that, because the
quasiparticle tunnel rates and their gradients have been approximated
as constants equal to unity for simplicity, the JQP curves in
figures~\ref{jqpc} and~\ref{tset} do not depend on the $V_{ds}$ bias choice.
With the dependences of the tunnel rates on the energies $E_{N+2,N+1}$
and $E_{N+1,N}$ properly taken into account [see equations
(\ref{tunnelrate}) and (\ref{dos})], one finds that the maximum and
minimum values of the $\mathcal{C}_1^{21}$
and $\mathcal{C}_2^{21}$ curves increase in magnitude as $V_{ds}$
decreases towards the onset for the JQP cycle at $e
V_{ds}=2\Delta+E_{c}$. In contrast, as we shall see below the DJQP curves
depend strongly on the
$V_{ds}$ bias choice even when the quasiparticle tunnel rates are
approximated as constants.

The results obtained from the numerical integrations of the master
equations agree well with those obtained within the Born-Markov
approximation. As expected, the agreement between the two improves
as the magnitude of $\kappa$ is reduced, in accord with our use of
the approximation $\kappa\ll 1$ in deriving the reduced master
equation.

In addition, the numerical integration of the master equations
gives us one more piece of information that we could not have
obtained from our calculations within the Born-Markov
approximation. Very close to the resonance we find that the
evolution of $\langle x(t)\rangle$ no longer matches that of a
damped harmonic oscillator.\footnote{The points plotted in
figure~\ref{jqpc} are all obtained from a fit to the behaviour of
a damped harmonic oscillator. Close to the resonance where this
fit is not readily obtained we were not able to extract values for
$\gamma_{\rm SET}$ and hence no corresponding data points are
plotted for $\mathcal{C}^2_{21}$}. Thus, it seems that very close
to the resonance the action of the SSET on the resonator is not
analogous to a thermal bath. Although we have not carried out a
systematic investigation of the region in which deviations from
thermal bath-like behaviour occur, we do find that the width of
the region (in terms of $E_{N+2,N}$) broadens with increasing
$\kappa$. This suggests that the weak coupling (Born)
approximation we use  to derive the Fokker-Planck equation may
break down for $m\omega_0^2x_s^2\gtrsim E_{N+2,N}$ (i.e., $\kappa
\gtrsim \tilde{E}_{N+2,N}$).

\subsection{DJQP resonance}

\begin{figure} \centering {\epsfig{file=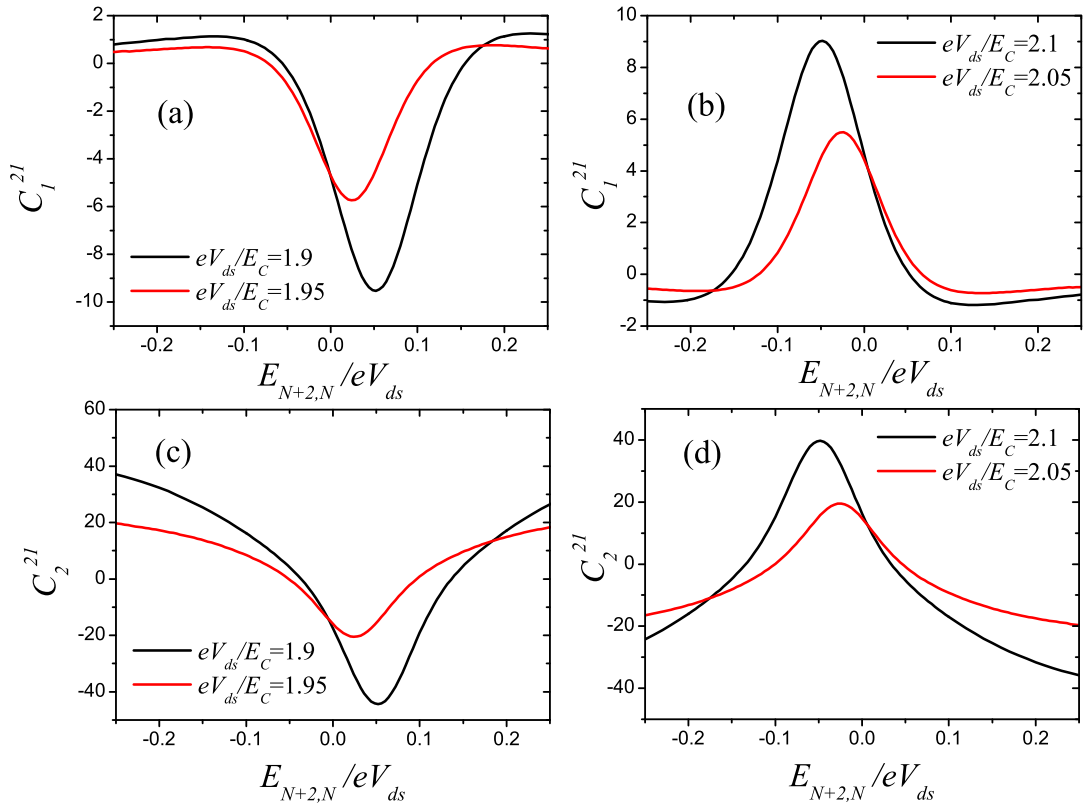,width=16cm}}
\caption{Numerical calculations of the coefficients
$\mathcal{C}_1^{21}$ and $\mathcal{C}_2^{21}$ in the vicinity of
the DJQP resonance.  The different values of $E_C/eV_{ds}$
correspond to points at different distances below, (a) and (c),
and above, (b) and (d), the centre of the DJQP resonance (see
figure \ref{map}). As before, we choose $\epsilon_J=1/16$, $r=1$
and set the quasiparticle tunnel rates and their gradients to
unity.} \label{djqp}
\end{figure}

\begin{figure}
\centering {\epsfig{file=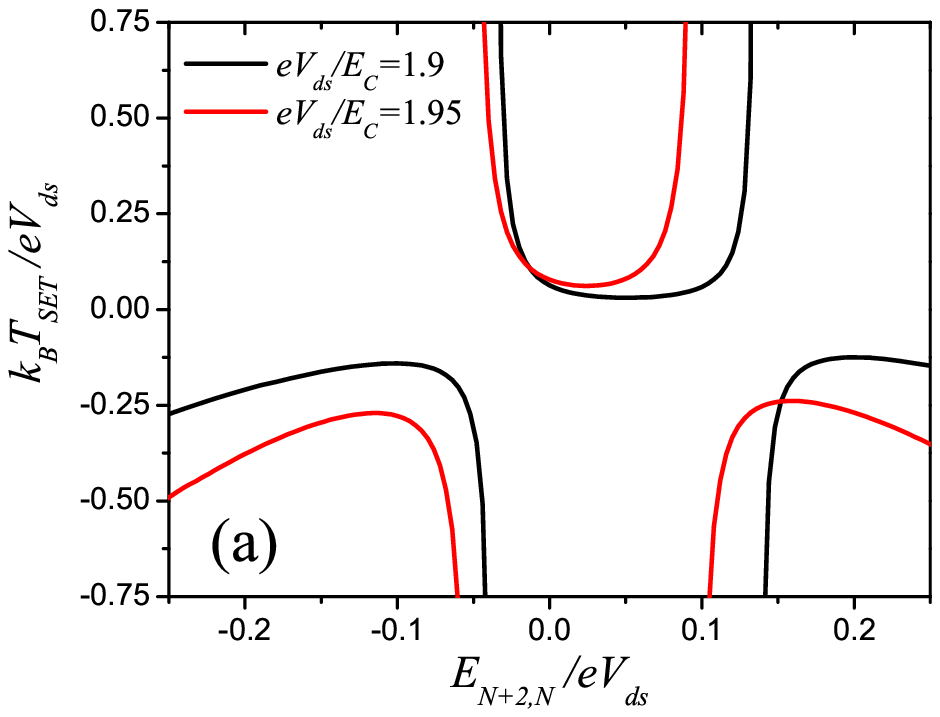,width=9cm}
\epsfig{file=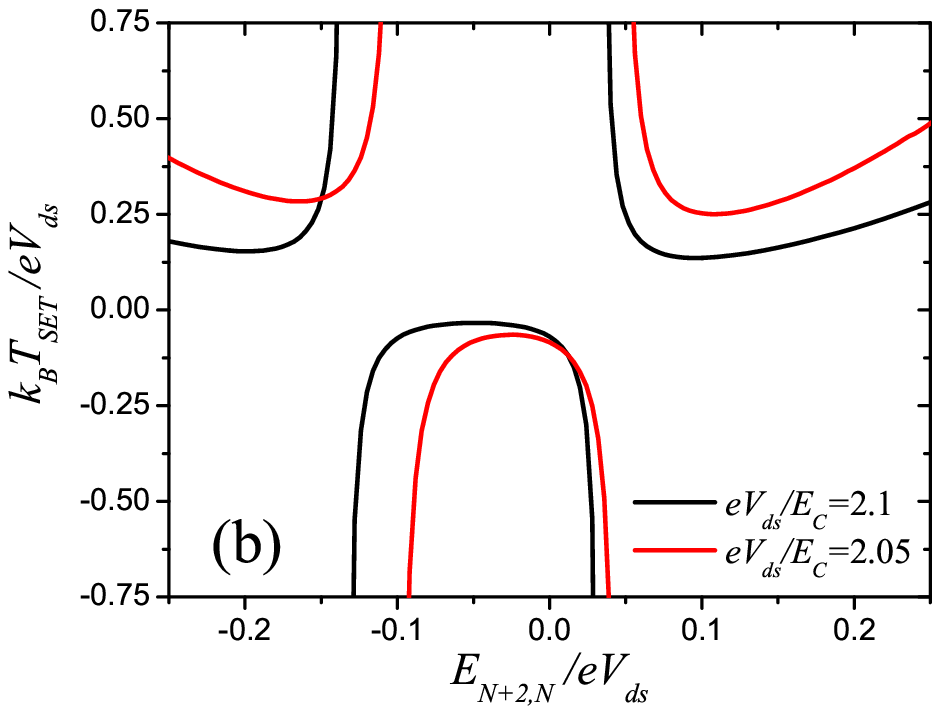,width=9cm}} \caption{Numerical calculation
of $T_{\rm SET}$ (a) below and (b) above the DJQP resonance. The
parameters are the same as those used for figure~\ref{djqp}.}
\label{tsetdjqp}
\end{figure}

The coefficients $\mathcal{C}_1^{21}$ and $\mathcal{C}_2^{21}$
calculated within the Born-Markov approximation are shown in
figure \ref{djqp}, while the corresponding variation in $T_{SET}$
is shown in figure \ref{tsetdjqp}. To facilitate comparisons with
the JQP resonance we have chosen $\tilde{E}_{N+2,N}$ and
$eV_{ds}/E_c$ as independent variables, in terms of which the
other energy detuning is
$\tilde{E}_{N+1,N-1}=\tilde{E}_{N+2,N}-4(E_c/eV_{ds})+2$. The
relation between the range of SET bias points covered in figures
\ref{djqp} and \ref{tsetdjqp} are illustrated schematically in
figure \ref{map}.

The DJQP resonance is much more complex than the JQP one and our
results by no means cover the whole range of the relevant
parameter space. However, by comparing figures \ref{jqpc} and
\ref{djqp}, it is clear that the magnitude of
$\mathcal{C}_2^{21}$, and hence $\gamma_{SET}$, is generally much
larger for the DJQP resonance than for the JQP one. As we shall
discuss below,
this implies
that in an experiment the effects of the back-action due to the
SSET on the resonator dynamics will be stronger close to the DJQP
resonance.

By analogy with the JQP resonance, it seems inevitable that there
will be regions very close to the DJQP resonances where our
Born-Markov approximation will fail to capture fully the physics
of the system. However, because of the additional complexity of
the DJQP cycle, we have not as yet performed the necessary
numerical integrations of the master equations required to
determine exactly where and how the Born-Markov approach breaks
down.

\section{Discussion and conclusions}

We now turn to consider the implications of our results for
experiments on nanomechanical-SET systems and to explore the
analogies which exist with other physical systems.

In an experiment the overall state of the resonator would be
determined by the combined effects of the SSET and the
external surroundings of the resonator other than the SSET,
which can be modelled by a damping rate,
$\gamma_e$, and a temperature, $T_e$. The overall effective
temperature and damping rate of the resonator would be given by
the weighted averages~\cite{armour},
\begin{eqnarray}
\gamma_{\rm eff}T_{\rm eff}&=&\gamma_{\rm SET} T_{\rm SET}+\gamma_{e}{T_e}\\
\gamma_{\rm eff}&=&\gamma_{\rm SET}+\gamma_e.
\end{eqnarray}
From these relations, it is clear that when $\gamma_{\rm SET}$
becomes negative $\gamma_{\rm eff}$ will decrease, eventually
leading to a dynamic instability in the state of the resonator
when $\gamma_{\rm eff}\leq 0$. However, we should point out that
the model described here is not sufficiently robust to explore
precisely what would happen in this regime.

The expressions for the effective temperature and damping rate also
make clear that  for $\gamma_{\rm SET}\gg \gamma_e$,
the effective temperature of
the resonator will be dominated by $T_{\rm SET}$. This has a number of
important implications given the minima in $T_{\rm SET}$ which develop
close to the JQP and DJQP resonances. For example, the minimum
value of $T_{\rm SET}$ which is simply proportional to the
quasiparticle decay rate can easily be as low as $\sim 100$mK,
suggesting that the electronic back-action on the resonator could
be demonstrated in a dramatic way by using the SSET to cool the
resonator when $T_e > T_{\rm SET}$. Indeed a close analogy can be made
between the temperature of a resonator coupled to a SSET and that
of (two-level) atoms undergoing Doppler cooling due to
counter-propagating laser beams~\cite{stenholm,lett}.

In Doppler cooling, the interaction between laser beams and two of
the atoms' energy levels leads to an effective damping of the
atomic translational motion. The atoms are cooled to a temperature
that depends on the detuning of the laser light from the atomic
resonance\footnote{Note we define the detuning, $\Delta$, with the
opposite sign to that given in Ref.~\cite{lett}.}, $\Delta$, and
the decay rate of the excited atomic state, $\Gamma_e$, given by
the relation~\cite{lett}
\begin{equation}
k_{\rm B}T_{\rm
Doppler}=\frac{\hbar}{4}\frac{\Gamma_e^2+4\Delta^2}{2\Delta}.
\end{equation}
This equation has an almost identical form to that for $T_{\rm
SET}$ close to the JQP resonance [equation (\ref{approxTSET})]. It
is interesting to note that a more direct analogy between Doppler
cooling and a system consisting of a resonator coupled to a Cooper
pair box (CPB)~\cite{ABS} addressed by an additional, fixed
voltage gate was suggested recently~\cite{martin}. The effective
temperature of the resonator in that case takes exactly the same
form as that near the JQP resonance when an appropriate ac-voltage
is applied to the extra fixed gate~\cite{martin}, although the
relevant decay rate is not related to quasiparticle tunnelling and
the associated expression for the damping of the resonator takes a
different form to that considered here.

Another system that might be expected to have very similar
dynamics to the SSET-resonator system is that of a double quantum
dot (DQD) gated by a mechanical resonator~\cite{brandes}. Since
DQDs display an electronic resonance which is in many ways
analogous to the JQP cycle, the dynamics of a resonator coupled to
a DQD should be very similar to that of one coupled to a SSET near
the JQP resonance. Experiments on DQD-resonator systems have not
yet been performed, but work on gating individual quantum dots by
mechanical resonators~\cite{susdot} demonstrates that such systems are
certainly a feasible prospect.

Our work raises a number of interesting questions for future
research. For example, in the regime that we examined, the
strongly non-linear properties of the quasiparticle tunnelling
rates played no role. It would be interesting to examine the
dynamics of the resonator for parameters close to the threshold
for quasiparticle tunnelling so that the motion of the resonator
itself could control whether or not tunnelling occurs. Also of
interest is our finding that the self-consistent Born-Markov
approximation method apparently cannot capture the physics of the
system when it is tuned extremely close to the JQP resonance; this
raises the intriguing possibility that the resonator dynamics in
this regime differs substantially from the thermalized dynamics
described here. Furthermore, we have not yet examined the
conditions under which Born-Markov approximation breaks down for
the DJQP resonance, though this is an area we plan to explore in
future work.

 In conclusion, we have found
that, like a normal state SET, a superconducting SET in the vicinity of either
the JQP or DJQP resonance acts on a nanomechanical resonator like
an equilibrium thermal bath for sufficiently weak
electro-mechanical coupling and for sufficiently large separation
between the electrical and mechanical time-scales. However, the
effective temperature, damping and frequency shift of the
resonator due to a SSET close to the JQP and DJQP resonances take
very different forms, both from each other and from the normal state
SET. In particular, the magnitude and even the sign of the
effective temperature and damping for the SSET depend very
sensitively on where it is tuned with respect to either the JQP
or DJQP resonance.

\ack

We gratefully acknowledge a number of very fruitful discussions
with O.\ Buu, M.D.\ LaHaye, A.J.\ Rimberg, K.C.\ Schwab and
especially A.A.\ Clerk. MPB would also like to thank  S.\ Tamura
and the University of Hokkaido for their hospitality where some of
this work was carried out, as well as the support of a Japan
Society for the Promotion of Science Fellowship. MPB acknowledges
funding from the NSF under NIRT grant CMS-0404031. JI and ADA
acknowledge support from the EPSRC under grant GR/S42415/01.

\appendix
\section{Master Equations for the DJQP Resonance}
In this Appendix we present the semiclassical master equations for
the SSET-resonator system in the vicinity of the DJQP resonance.
As shown schematically in figure~\ref{JQP}b, the DJQP
cycle~\cite{djqp1,djqp2,clerk4} involves resonant Cooper pair
tunnelling at each junction alternating in turn with two
quasiparticle tunnelling events.

The semiclassical master equations for the SSET-resonator near
the DJQP resonance are derived using the same procedure as for the
JQP resonance. Written in dimensionless notation, they take the
form~\cite{clerk4},
\begin{eqnarray}
\fl \dot{\rho}_{N-1}= \epsilon_{\rm HO}^2[x+(N-1)]\frac{\partial
\rho_{N-1}}{\partial v}- v\frac{\partial \rho_{N-1}}{\partial
x}+i\pi\epsilon_J\left(\rho_{N+1,N-1}-\rho_{N-1,N+1}
\right)\nonumber\\
\fl -\left[\tilde{\Gamma}(\tilde{E}_{N-1,N})+
\tilde{\Gamma}'(\tilde{E}_{N-1,N})\kappa x\right]\rho_{N-1}\\
\fl \dot{\rho}_{N}= \epsilon_{\rm HO}^2(x+N)\frac{\partial
\rho_{N}}{\partial v}-
v\frac{\partial \rho_{N}}{\partial x}+i\pi\epsilon_J
\left(\rho_{N+2,N}-\rho_{N,N+2}\right)\nonumber\\
\fl +\left[\tilde{\Gamma}(\tilde{E}_{N-1,N})+\tilde{\Gamma}'
(\tilde{E}_{N-1,N})\kappa x\right]\rho_{N-1}\\
\fl \dot{\rho}_{N+1}= \epsilon_{\rm HO}^2[x+(N+1)]\frac{\partial
\rho_{N+1}}{\partial v}- v\frac{\partial \rho_{N+1}}{\partial
x}-i\pi\epsilon_J\left(\rho_{N+1,N-1}-\rho_{N-1,N+1}
\right)\nonumber\\
\fl +\left[\tilde{\Gamma}(\tilde{E}_{N+2,N+1})+\tilde{\Gamma}'
(\tilde{E}_{N+2,N+1})\kappa x\right]\rho_{N+2}\\
\fl \dot{\rho}_{N+2}= \epsilon_{\rm HO}^2[x+(N+2)]\frac{\partial
\rho_{N+2}}{\partial v}- v\frac{\partial \rho_{N+2}}{\partial
x}-i\pi\epsilon_J\left(\rho_{N+2,N}-\rho_{N,N+2}
\right)\nonumber\\
\fl -\left[\tilde{\Gamma}(\tilde{E}_{N+2,N+1})+\tilde{\Gamma}'
(\tilde{E}_{N+2,N+1})\kappa x\right]\rho_{N+2}\\
\fl \dot{\rho}_{N-1,N+1}=\epsilon_{\rm HO}^2(x+N)\frac{\partial
\rho_{N-1,N+1}}{\partial v}- v\frac{\partial \rho_{N-1,N+1}}{\partial
x}+i\pi\epsilon_J\left(\rho_{N+1}-\rho_{N-1}\right)\nonumber\\
\fl +\left[2\pi i r\left(\tilde{E}_{N+1,N-1}+2\kappa
x\right)-\frac{1}{2}\left(\tilde{\Gamma}(\tilde{E}_{N-1,N})+\tilde{\Gamma}'
(\tilde{E}_{N-1,N})\kappa
x\right)\right]\rho_{N-1,N+1}\\
\fl \dot{\rho}_{N,N+2}=\epsilon_{\rm HO}^2[x+(N+1)]\frac{\partial
\rho_{N,N+2}}{\partial v}- v\frac{\partial \rho_{N,N+2}}{\partial
x}+i\pi\epsilon_J\left(\rho_{N+2}-\rho_{N}\right)\nonumber\\
\fl +\left[2\pi i r\left(\tilde{E}_{N+2,N}+2\kappa
x\right)-\frac{1}{2}\left(\tilde{\Gamma}(\tilde{E}_{N+2,N+1})
+\tilde{\Gamma}'(\tilde{E}_{N+2,N+1})\kappa
x\right)\right]\rho_{N,N+2}.
\end{eqnarray}
The relevant energy differences for the quasiparticle and Cooper
pair tunnelling processes involved in the DJQP cycle are given by
\begin{eqnarray}
E_{N+2,N+1}&=&-2E_c(N_g-N-3/2)+eV_{ds}\\
E_{N-1,N}&=&2E_c(N_g-N+1/2)
\end{eqnarray}
and
\begin{eqnarray}
E_{N+2,N}&=&-4E_c(N_g-N-1)\\
E_{N+1,N-1}&=&-4E_c(N_g-N)+2eV_{ds},
\end{eqnarray}
respectively. The DJQP resonance occurs when
$E_{N+2,N}=E_{N-1,N+1}=0$, i.e.\ when $eV_{ds}=2E_c$ and
$N_g=N+1$. Notice that the cycle also requires that the
quasiparticle tunnelling processes are not energetically
forbidden, which implies that $E_c> 2\Delta/3$, if we assume for
simplicity that the superconductors are at zero temperature.

These master equations give rise to a Fokker-Planck equation with
the same form as Equation (\ref{fokkerplanck}), but with different
expressions for the coefficients ${\mathcal{C}}_1^{21}$,
${\mathcal{C}}_2^{21}$, ${\mathcal{C}}_1^{22}$ and
${\mathcal{C}}_2^{22}$. These coefficients can be calculated
numerically, in the same way as those for the JQP resonance, to
give the results presented in section 3.3.

\end{document}